\pdfoutput=1
\makeatletter
\let\blx@rerun@biber\relax
\makeatother

\documentclass{bvm} 

\begin{document}

\newcommand{\bvmyear}{2023}

\selectlanguage{english} 

\title{Unsupervised detection of small hyperreflective features in ultrahigh resolution optical coherence tomography}

\titlerunning{Detection of hyperreflective features in UHR OCT}

\author{
	Marcel \lname{Reimann} \inst{1,2}, 
	Jungeun \lname{Won} \inst{1},
	Hiroyuki \lname{Takahashi} \inst{1,3},
	Antonio \lname{Yaghy} \inst{3},
	Yunchan \lname{Hwang} \inst{1},
	Stefan \lname{Ploner} \inst{1,2},
	Junhong \lname{Lin} \inst{1},
        Jessica \lname{Girgis} \inst{3},
        Kenneth \lname{Lam} \inst{3},
	Siyu \lname{Chen} \inst{1},
	Nadia~K. \lname{Waheed} \inst{3},
	Andreas \lname{Maier} \inst{2}, 
	James~G. \lname{Fujimoto} \inst{1}
}

\authorrunning{Reimann et al.}

\institute{
\inst{1} Department of Electrical Engineering and Computer Science, Research Laboratory of Electronics, Massachusetts Institute of Technology, USA\\
\inst{2} Pattern Recognition Lab, Friedrich-Alexander-Universität Erlangen-Nürnberg, Germany\\
\inst{3} New England Eye Center, Tufts Medical Center, USA\\
}

\email{marcel.reimann@fau.de}

\maketitle

\begin{abstract}
Recent advances in optical coherence tomography such as the development of high speed ultrahigh resolution scanners and corresponding signal processing techniques may reveal new potential biomarkers in retinal diseases. Newly visible features are, for example, small hyperreflective specks in age-related macular degeneration. Identifying these new markers is crucial to investigate potential association with disease progression and treatment outcomes. Therefore, it is necessary to reliably detect these features in 3D volumetric scans. Because manual labeling of entire volumes is infeasible a need for automatic detection arises. Labeled datasets are often not publicly available and there are usually large variations in scan protocols and scanner types. Thus, this work focuses on an unsupervised approach that is based on local peak-detection and random walker segmentation to detect small features on each B-scan of the volume. 
\end{abstract}

\section{Introduction}
Age-related macular degeneration (AMD) ranks fourth on the leading causes of vision loss worldwide~\cite{3351-01}. To develop new potential treatments, trial endpoints need to be defined based on early markers of disease progression. New systems such as the ultrahigh resolution spectral domain OCT (UHR SD-OCT) enable the detection of smaller hyperreflective features in the human retina, such as specks~(HRS), which have been suggested as indicators for cellular activities associated with visual dysfunction in AMD~\cite{3351-01}.

For the detection of conventional, but larger, hyperreflective foci (HRF) multiple algorithms already exist~\cite{3351-03, 3351-06,3351-07,3351-08}. They can be divided into algorithms that work on volumes, B-scans or on enface projections of specific retinal layers. Most approaches focus on intensity based thresholds applied to 8-bit images as they are available in commercial OCT instruments. More recent methods utilize optical attenuation coefficients or deep learning. However, most of these algorithms do not utilize the full dynamic range of the OCT signal, suffer from limited resolution scanners, or do not consider the limited amount of available labeled data.~\cite{3351-03, 3351-06,3351-07,3351-08}

To overcome these issues, we propose an unsupervised algorithm that is able to detect features in anisotropic high-definition volumetric OCT datasets by applying a combination of out-of-the-box image analysis tools. In addition, we show that the algorithm is applicable to different scan patterns, including anisotropic high-definition scans and isotropic motion corrected data from multiple merged scans. 

\section{Materials and Methods}
\subsection{Data acquisition and preprocessing} 
\label{3351-sec-data}
Studies were performed under an IRB approved protocol at MIT and the NEEC. Written informed consent was obtained. Data was collected by a UHR SD-OCT prototype with an axial resolution of \textasciitilde 2.7\ts\textmu m. The anisotropic high-definition volumetric scan was acquired over a $9 \times 6$\ts mm field with an A-scan spacing of~5\ts \textmu m and a B-scan spacing of~25\ts \textmu m. No B-scan or volumetric averaging was performed and the data was stored using 32~bit floating point values. After resampling, the pixel size is 0.89 \ts \textmu m in axial direction and the B-scan dimensions are $1600 \times 1800$ pixels. 

For evaluation of the algorithm,
a total of 49~B-scans from 10 different eyes diagnosed with early~(3) and intermediate~(7)~AMD were selected based on potential appearences of HRS/HRF. Three expert readers independently labeled HRF and HRS and their results were combined using STAPLE~\cite{3351-05} with a threshold of $\frac{2}{3}$ and used as ground truth. Of all annotated feature instances 32.2\ts\% were labeled by all readers. After STAPLE was applied, 49.24\ts\% of all previously labeled feature instances remained as ground truth. 

To test the applicability of the algorithm on different imaging protocols, on the same device, 4 isotropic scans with $500 \times 500$ A-scans were acquired over a $6 \times 6$~mm field with alternating horizontal and vertical B-scan directions. The transverse and (resampled) axial pixel spacings were 12\ts \textmu m and 1.78\ts \textmu m, respectively. The scans were motion corrected and merged using the approach of Ploner~et~al.~\cite{3351-02}

\subsection{Algorithm description}
\label{3351-sec-algo}
The algorithm is separately applied to each B-scan of the volumetric scan. 
An essential requirement is the availability of retinal layer segmentation for the posterior of the outer plexiform layer (OPL), external limiting membrane (ELM) and the inner segment/outer segment border (IS/OS), as detecting HRF and HRS in the outer nuclear layer and external to ELM was our priority. For simplification, we refer to this combined region as ONL. 
As a first step, the layer boundaries are smoothed and the ONL mask eroded to account for possible inaccuracies in the provided layer segmentation. Next, varying illumination (OCT signal) within the ONL was equalized by dividing with a bias field. The field was computed using the axial mean projection of the ONL without considering zero values, and then smoothed by a 30 pixels standard deviation Gaussian kernel following the implementation by Kraus~et~al.~\cite{3351-04} To prevent false positives on the highly reflective ELM,  each pixel value in the area of 5 pixels above and below was adjusted so that their axial mean is just below the axial mean of the ONL. For the detection of high intensity features, we utilize local maxima detection. Very isolated maxima are deleted using a hit-or-miss transform with a single pixel structuring element to prevent false positives. After defining the detected local maxima as foreground pixels, the gray scale values of the B-scan are re-scaled to a range of [-1,1], based on the minimum and maximum values of the B-scan, and background pixels are defined to be below a threshold of -0.9. Random walker segmentation is then applied to expand the regions around the local maxima~\cite{3351-09}.

In the end, multiple thresholds are applied to eliminate false positives by constraining the results to optical properties of the system and clinically relevant features. The minimum size of the detected features is set to be at least 4 pixels in axial dimension and 3 pixels transverse, corresponding to laser beam spot size and optical resolution. The pixel intensities of the feature are compared to a threshold derived from all ONL values covering the same transverse span. The threshold requires the maximum intensity to be larger than the mean plus two times the standard deviation of the ONL values. In addition, the axial mean position of the feature should be within a relative position of 0.25 to 0.8 between the IS/OS (0.0) and OPL (1.0). If the positional requirement is not fulfilled, but the maximum intensity of the detected feature is twice as high as the mentioned threshold, the feature is still considered as detected in the evaluation. Another hard requirement is that the features must be detached from the IS/OS and OPL to be considered as detection, as defined by a previous study by Echols~et~al.~\cite{3351-01} All of these constraints were also applied to the ground truth after labeling.

\section{Results}
\label{3351-sec-results}
\begin{figure}[b]
	\setlength{\figwidth}{\textwidth}
	\centering
	\includegraphics[width=\figwidth]{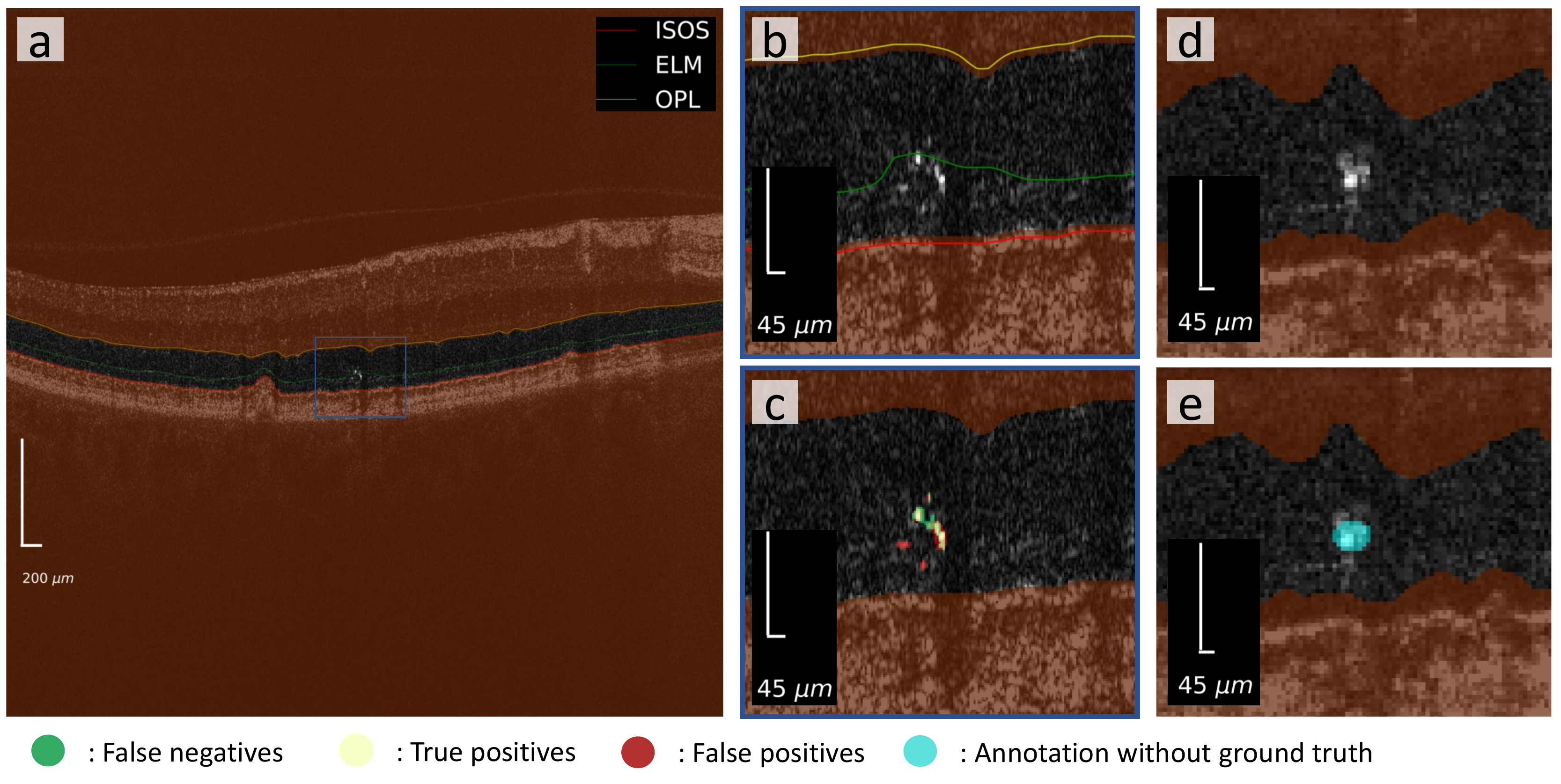}
	\caption{a) Annotated B-scan from $9 \times 6$~mm volume including layer segmentation and algorithm results b)-c) Zoomed in version of a reference without annotations and including annotations by algorithm and expert reader d)-e) Zoomed in version of a corresponding frame in the merged $6 \times 6$~mm volume without and with algorithm annotations.}
	\label{3351-fig-01}
\end{figure}
\begin{figure}[b]
	\setlength{\figwidth}{\textwidth}
	\centering
	\includegraphics[width=\figwidth]{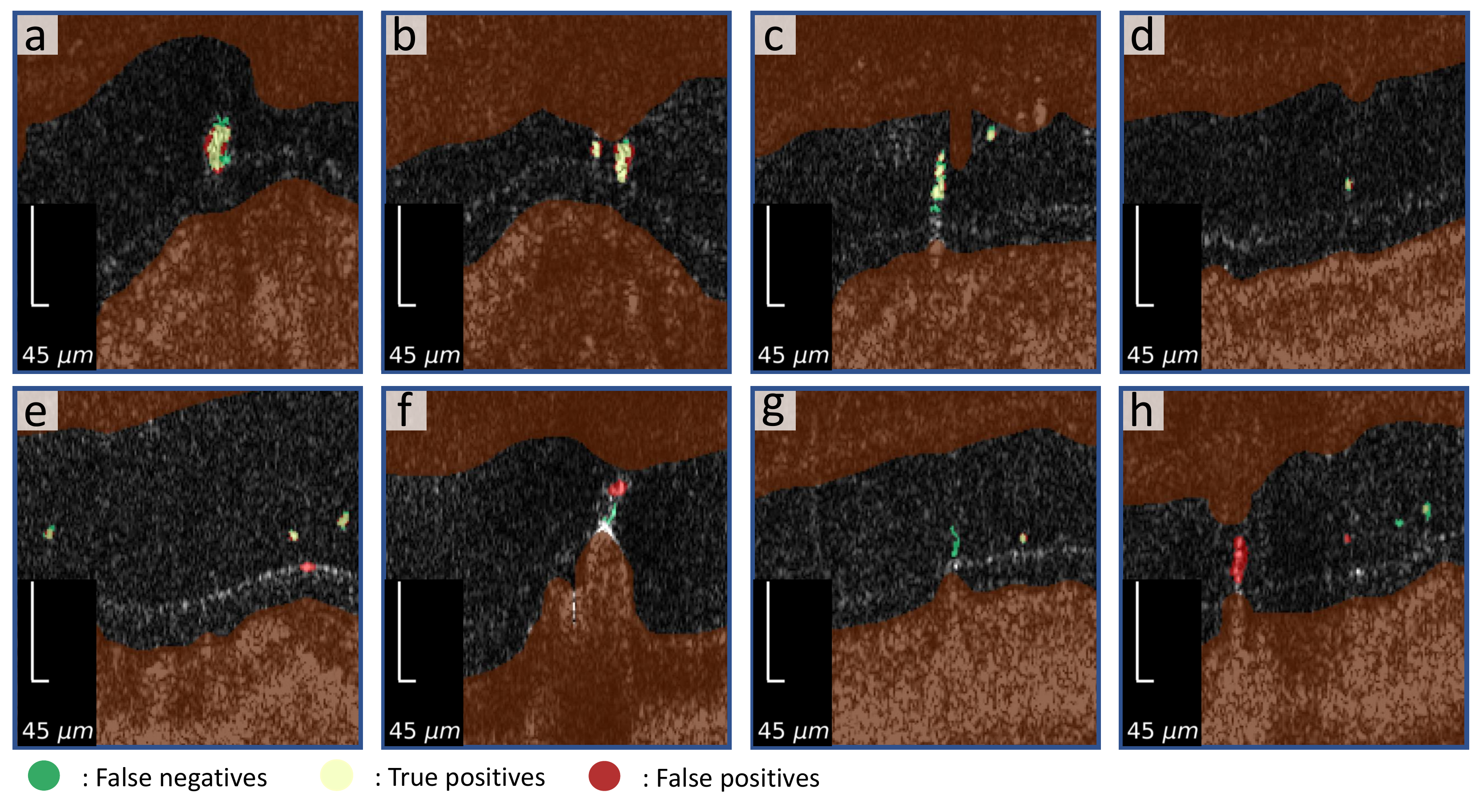}
	\caption{a)-d) Example images of well segmented hyperreflective features. e) False positive on ELM with high intensity. f) Ground truth annotations in contact with the OPL and algorithm annotations in contact with the IS/OS are, because of the defined constraints, deleted and not shown. Inaccurate layer segmentation is also visible. g) False negative feature appears clearly in neighboring scan, but has low intensity in the current B-scan. h) Ground truth in contact with IS/OS and, therefore, not considered or shown.}
	\label{3351-fig-02}
\end{figure}
The evaluation is based on binary classification metrics, in which labeled hyperreflective features were considered positives. Because only very few pixels per B-scan were labeled as positives we chose metrics that take high class imbalances into account. Precision and recall were calculated based on accumulated confusion matrix values over all B-scans. Results are shown in table~\ref{3351-tab-evaluation} and figure~\ref{3351-fig-01}. The per instance results are calculated based on 66 true positives out of 93 labeled features and 44 false positives. A detection was considered as true positive, if there is at least one pixel overlap with an area in the ground truth mask. Also shown in table~\ref{3351-tab-evaluation} are results reported by Schlegl~et~al.~\cite{3351-03} for the detection of large conventional HRF. Their deep learning method was trained on 1051 B-scans acquired on Cirrus HD-OCT and Spectralis OCT and evaluated for Cirrus and Spectralis images separately. They include cases with AMD, diabetic macular edema and retinal vein occlusion. They harmonized the different scan protocols by re-sampling the images achieving unified pixel dimensions of \textasciitilde6\ts \textmu m axial and \textasciitilde11.5\ts \textmu m transverse.  

In our case, the average dimensions of the labeled features are 5.0 pixels in transverse dimension and 9.8 pixels axial corresponding to 25\ts \textmu m transverse and \textasciitilde 8.7\ts \textmu m axial based on the pixel spacing. The features cover an average area of 36 pixels or \textasciitilde 160.2\ts \textmu m$^2$. Unfortunately, Schlegl et al. do not provide average sizes of their features. However, on their figures, they appear to be much larger, corresponding to conventional HRF. In addition, they also include features in different retinal layers. Due to the difference in sizes and scale, the performance of their algorithm is not directly comparable to our results, yet their results show the performance of a different algorithm on a similar task.  
\begin{table}[tb]
	\centering
	\caption{Per pixel and per instance evaluation for the detection of small hyperreflective specks and foci in comparison to pixel-wise results for larger conventional hyperreflective foci reported by Schlegl et al.~\cite{3351-03} for data acquired on a Cirrus/Spectralis OCT scanner.}
	\label{3351-tab-evaluation}
	\begin{tabular*}{\textwidth}{l@{\extracolsep\fill}ccc}
	\hline
          & Evaluation per pixel & Evaluation per instance & Schlegl et al. per pixel \\ \hline
         
Precision & 49.00 \%           & 60.00 \%               & 66.55 \% / 55.98 \%      \\
Recall    & 63.25 \%           & 70.97 \%               & 64.01 \% / 73.32 \%      \\
F1 score  & 55.22 \%           &  65.03\%               & 65.26 \% / 63.49 \%   
\\ \hline
\end{tabular*}
\end{table}
A more detailed comparison of our algorithm and the ground truth yields the following observations~(\figurename~\ref{3351-fig-02}). Labeled features with slightly higher intensity values than the background are difficult to detect and the reader agreement on these is not high. This also applies to boundary regions of the features because of the smooth transition to background intensities and a missing consensus of where to delineate feature boundaries. Another situation of disagreement can occur on much higher intensity values on the ELM compared to the rest of the ELM. They are usually detected by the algorithm, but usually not annotated by the human experts. In addition, for features that are close to IS/OS and OPL, it is sometimes unclear, if they are detached from these layers. It occurs that only either the algorithm or the readers include or exclude them, which has a high impact on the scores as the regions are comparably large. Also, features that clearly appear in neighboring B-scans, but are not clearly visible in the current frame, are occasionally annotated by human readers and not by the algorithm.

In figure~\ref{3351-fig-01} we show, that the algorithm is also applicable to data acquired with a different scan protocol and averaged scans. The algorithm only requires the adjustment of hyperparameters that can easily be converted based on the pixel sizes. These include the illumination correction and the size thresholds.

\section{Discussion}
Compared to previous work, this algorithm has the advantage that it is inherently explainable and does not require labeled data. For the deployment on new datasets only certain parameters have to be converted to match the pixel spacing of the new input. We demonstrated the generalization to data that was not used during the development and hyperparameter selection. Although the study data can, at this point, not be made publicly available, the method can readily be re-implemented and used out of the box. Another advantage of the algorithm is that it utilizes depth information, which is not possible for algorithms that only work on enface projections of the volume. By this, we exploit one of the big strengths of our prototype, namely the high resolution. 

We choose thresholds based on the optical properties of the OCT system and include the smallest features that are still possible to detect. As a result, this has a great negative influence on the evaluation scores, because the features are relatively small and single-pixel misclassifications have a large impact on pixel-wise metrics. However, it is not proven that the exact size of the feature is of clinical relevance. In contrast, studies suggest that the number of detected HRF and HRS might be of clinical relevance and can be associated with delayed rod-mediated dark adaptation, which is linked to AMD progression~\cite{3351-01}. Therefore, we propose to rather look at detected feature instances instead of semantic segmentation, which is a measure used by Echols~et~al.~\cite{3351-01} Comparing the algorithm to other state-of-the-art methods is very difficult and we only report the results of Schlegl et al.~\cite{3351-03} to give an idea about the performance of most recent methods on similar tasks. However, most of the datasets are collected with different devices and scan protocols and differ in resolution, dynamic range and image sizes. For example, Schlegl et al. cannot report on features as small as ours because of a lower resolution. In addition, there is not yet a clinical consensus on what HRS exactly are and how to label them. HRS sizes are usually below or close to the resolution of commercial systems and have not been investigated on a larger scale. 
Our work will provide a tool for larger clinical studies on HRF and HRS distributions close to optical resolution and can  potentially be used on datasets collected at different sites with different devices. The only requirement is the availability of layer segmentation for the IS/OS, OPL and ELM. However, the ELM scaling is an optional step in the processing and can be omitted. In the future, the algorithm could be adapted to other retinal layers. For now, the focus lies on the ONL, as we expect more HRF and HRS in this layer during the early stages of AMD. In addition, an extension for isotropic volumes should be considered that operates in 3D to, for example, overcome the issue of detached or non-detached features from the ONL mask boundaries. At the time of submission only limited data for isotropic volumes was available. More clinical studies with larger patient cohorts will follow.
\begin{acknowledgement}
	We acknowledge funding by the National Institutes of Health, project 5-R01-EY011289-36, and the German Research Foundation, project 508075009.
\end{acknowledgement}

\printbibliography

\end{document}